\documentclass[11pt]{article}

\usepackage[margin=2.5cm]{geometry}
\usepackage{amsmath,amssymb,amsfonts}
\usepackage{graphicx}
\usepackage{booktabs}
\usepackage[super,square,sort&compress]{natbib}
\usepackage[colorlinks=true,linkcolor=blue,citecolor=blue,urlcolor=blue]{hyperref}
\usepackage{lineno}
\usepackage{setspace}
\usepackage{authblk}

\usepackage{titlesec}
\titleformat{\section}{\large\bfseries}{}{0pt}{}
\titleformat{\subsection}{\normalsize\bfseries}{}{0pt}{}
\titleformat{\subsubsection}{\normalsize\itshape}{}{0pt}{}

\newcommand{\kB}{k_{\mathrm{B}}}
\newcommand{\CSO}{C_{\mathrm{SO}}}
\newcommand{\zeff}{z_{\mathrm{eff}}}
\newcommand{\cL}{c_{\mathrm{L}}}

\title{\bfseries One geometric barrier unifies melting, vitrification and
jamming of hard spheres in all dimensions}

\author[1]{Sujin B.\ Babu}
\affil[1]{Out of Equilibrium Group, Department of Physics, Indian Institute of Technology Delhi, Hauz Khas, New Delhi 110016, India}

\date{}

\begin{document}

\maketitle

\begin{abstract}
\noindent\textbf{%
The Lindemann criterion that a solid loses stability once atomic vibrations reach roughly a tenth of the interparticle spacing, has remained an empirical rule for over a century. The numerical value was reproduced by mode-coupling and replica theories but never isolated as the consequence of a simple, verifiable argument. Here we show that for hard spheres in $d$ dimensions the rule follows from three exact geometric ingredients. The contact theorem fixing the coordination number from the equation of state, an isotropy identity fixing how non touching neighbors project onto an escape direction, and a first-passage argument which is derived, in which the elementary hop spans one interparticle spacing rather than one particle diameter. The resulting parameter-free master equation locates the kinetic glass transition, random close packing, the Kauzmann point, glass close packing, and equilibrium crystal melting in $d=3$--$12$, each to within a few per cent of reported independent simulation and replica-theory values, and places all five on a single barrier surface. The theory makes two predictions that are verifiable, the Lindemann constant, $\cL(3)=0.13$ per neighbor spacing in $3$ dimensions derived from the theory, which must fall systematically with increasing dimensions. The other being in two dimensions, the current theory predicts the arrest in the volume fraction $\eta_g=0.781$, the jamming at $\eta=0.832$, and both steps of the two-stage melting scenario, all of which are already corroborated by independent simulations and experiments.}
\end{abstract}


Packings held together by many weak, distributed constraints rather than one strong bond pervade condensed matter, from atomic crystals and colloidal glasses to granular media and jammed emulsions. The Lindemann criterion~\cite{Lindemann1910} is the oldest quantitative statement of when such confinement fails. Accordingly a solid becomes unstable once the thermal excursions of its constituents exceed a fraction $\cL$ of the spacing between them. The value  $\cL\approx0.1$ across chemically unrelated systems has been numerically reproduced by mode-coupling theory~\cite{Bengtzelius1984} and by mean-field replica theory~\cite{ParisiZamponi2010}. These empirical correlations between vibrational amplitude and structural relaxation extend far beyond simple solids, extending to polymers and molecular glass formers~\cite{Larini2008,Ottochian2011}. However, in none of these frameworks is the number $0.1$ isolated as the output of an identifiable minimal calculation. A century after its formulation, the most widely used stability criterion in condensed matter remains a correlation which lacks derivation.

Hard spheres offer the one system in which the relevant geometry can be worked out exactly as the no interaction potential is simple, no closure approximation etc. Here we derive the Lindemann criterion for hard spheres in $d$ dimension, using only the equation of state and the kissing number of the local packing as input. The same relation then places several hallmarks of the hard sphere phase diagram that are normally treated by separate theories, equilibrium crystal melting, the kinetic glass transition, random close packing (RCP), the Kauzmann point and glass close packing (GCP), on a single geometric barrier surface, with each landmark occupying an essentially dimension-independent coordinate. Contrary to the present practice, the two steps taken as physical hypotheses are here supplied with derivations. The escape length that sets the barrier, and the identification of the barrier as a genuine relaxation exponent, both follow from a solved first-passage problem (Methods).

Crucially, the theory is not merely descriptive, it is verifiable on two independent fronts, and on both it has already predicted the correct results. First, evaluated on the crystal branch it returns the empirical hard-sphere melting amplitude, $\cL(3)=0.130$ per neighbor spacing, with no adjustment, and predicts that this classical constant is a specific to three-dimensional system that must decay monotonically with dimension, a statement directly testable in high-dimensional crystal simulations. Second, extended to two dimensions, it predicts dynamical arrest at $\eta_g=0.781$ and jamming at $\eta=0.832$, independent bidisperse hard-disk simulations~\cite{deGraaf2024}, quasi-two-dimensional colloid experiments~\cite{Lozano2019} and athermal jamming protocols~\cite{OHern2003} place these transitions at $0.776-0.79$ and $0.83-0.84$ respectively. Evaluated on its fluid and lattice branches, the same equation further locates both steps of the two-dimensional melting scenario, the liquid--hexatic and hexatic--solid transitions of hard disks~\cite{BernardKrauth2011} to within one per cent. The oldest rule of thumb in condensed matter is thus recast as a predictive, parameter-free theory.

\section{Results}

\subsection{The cage, made exact}

The argument has four physical ingredients, the cage a particle sits in is harmonic, its stiffness is fixed by how many neighbors touch it versus how many merely surround it, the non-touching neighbors are weighted by a purely geometric projection factor and the particle escapes only once it has moved roughly one neighbor spacing, not one particle diameter. Each ingredient is exact for hard spheres, in the sense that it introduces no approximation beyond the equation of state itself.

\textit{The cage is harmonic.} Hard spheres carry no potential energy, so the confinement a tagged particle feels is a statement about geometry and more arrangements of the surrounding particles exist for a centered configuration than for an off-centre one. This defines an entropic potential $\phi(u)=-\kB T\ln P(u)$. Because the packing is statistically isotropic, the cage centre is a stationary point of $\phi$, and the leading behaviour is quadratic, $\phi(u)\approx\tfrac{1}{2}\kappa u^{2}$. This is not an assumption about detailed structure it follows from symmetry and it is self-consistent: the resulting amplitude is about a tenth of the spacing, so the neglected anharmonic terms are smaller by that fraction squared (Methods). Equipartition then gives $\langle u^2\rangle=\kB T/\kappa$ per Cartesian direction and all the remaining physics is in $\kappa$.

\textit{Counting what touches the particle.} A tagged sphere's stiffness comes from resisting displacements that carry it into its neighbors. Some neighbors are in direct contact and resist immediately and others sit within the coordination shell without touching, and resist only to the extent that the motion closes the gap. The total number of neighbors within one diameter follows from the pressure via the exact contact theorem,
\begin{equation}
X = 2d\,(Z-1),
\label{eq:contact}
\end{equation}
where $Z(\eta,d)$ is the compressibility factor (Methods). A microscopic coordination number is thereby read directly from a macroscopic equation of state.

\textit{Why distant neighbors contribute less.} A neighbor off the escape axis feels only the projection of the tagged particle's motion onto its own line of sight. Averaged over an isotropic packing this projection is fixed by symmetry,
\begin{equation}
\bigl\langle(\hat e\cdot\hat n)^2\bigr\rangle = \frac{1}{d},
\label{eq:isotropy}
\end{equation}
with no dependence on density or structure (Methods). Weighting the $X-k_d$ non-touching neighbors ($k_d$ the kissing number) by this factor gives effective coordination $\zeff=k_d+(X-k_d)/d$ and, in the spirit of nonaffine elasticity~\cite{Wyart2005,Zaccone2011}, a stiffness factor 
\begin{equation}
\CSO(\eta,d) = 1-\frac{2d}{\zeff}.
\label{eq:cso}
\end{equation}
This vanishes at isostaticity ($X=k_d=2d$), rises towards unity when distant neighbors dominate, and turns negative for sub-isostatic lattices such as diamond. The cage spring constant is $\kappa\sigma^2/\kB T=2d\,\CSO(\eta,d)\,[Z(\eta,d)-1]$.

\textit{How far a particle must move to escape.} A cage fails once the tagged particle has swapped neighbors, which requires moving of order the centre-to-centre spacing, not the particle diameter:
\begin{equation}
\ell(\eta,d)\equiv\frac{L}{\sigma}
   =\Bigl(\frac{V_d}{2^d\eta}\Bigr)^{1/d},
\qquad
V_d=\frac{\pi^{d/2}}{\Gamma(d/2+1)}.
\label{eq:ell}
\end{equation}
At three-dimensional densities this is close to one diameter, which is why we think the two lengths have not been distinguished before, at higher $d$ the packing dilutes and $L$ falls well below $\sigma$ (Methods). Projecting the random escape hop by equation~(\ref{eq:isotropy}), the work against the cage in units of $\kB T$ is the Kramers exponent controlling $\tau_\alpha\propto e^{B}$:
\begin{equation}
B(\eta,d)=\ell(\eta,d)^{2}\,\CSO(\eta,d)\,[Z(\eta,d)-1],
\label{eq:master}
\end{equation}
and the Lindemann constant is $\cL=\ell/\sqrt{2dB}$. The explicit factor of $d$ in the stiffness cancels the $1/d$ from the projected hop at every dimension (Methods). Equation~(\ref{eq:master}), together with the threshold value of $B$, constitutes the complete theory.

Three distinct lengths must be kept apart. The cage-to-cage spacing $\ell$ is the distance between the centers of neighboring cages. The localization length $r_s=\sqrt{\langle u^2\rangle/2}$ is the amplitude entering the Debye--Waller factor, or equivalently the non-ergodicity parameter $f_q=\exp(-\tfrac{1}{2}q^2\langle u^2\rangle)$, with $q$ the magnitude of the wavevector. The projected escape length $a=\ell/\sqrt{d}$ along the direction of propagation shrinks faster than $\ell$ with increasing dimension and therefore controls the barrier and it varies from $a=0.56$ at $d=3$ to $a=0.28$ at $d=9$.

\subsection{The barrier is derived, not postulated}

The identification of $\ell$ as the escape length, and of $B$ as a genuine relaxation exponent, is not assumed here but follows from a first-passage calculation (Methods). The tagged particle diffuses in the harmonic cage until it reaches the projected escape distance $a=\ell/\sqrt{d}$ because $U(a)=B$ exactly, the mean first-passage time evaluates to
\begin{equation}
\ln\frac{\tau_\alpha}{\tau_c}
  = B-\tfrac{1}{2}\ln B+\ln N_{\mathrm{eff}},
\label{eq:mfpt}
\end{equation}
with $\tau_c$ the collision time and $N_{\mathrm{eff}}=(a/\lambda)^2$ the squared number of mean free path steps needed to traverse $a$, the mean free path $\lambda$ being fixed by the equation of state alone. Thus $B$ is the exact extensive term of a solved stochastic process, and the corrections grow only logarithmically. At the colloidal dynamical glass point $\eta_g(d{=}3)=0.571$ the extensive barrier is $B=17.14$ and the assembled first-passage log-time is $18.05$, compared with $B^{*}=18.42$ (that is, $\tau_\alpha/\tau_0=10^{8}$), the experimental definition of the glass transition in $d=3$. The single calibration of the theory is therefore itself computed it is the prefactor of equation~(\ref{eq:mfpt}), not an independent free parameter.

\subsection{Glass-transition densities across dimensions}

Averaging $B$ over the dynamical glass-transition densities reported by simulation in $d=3$--$12$ (ref.~\citenum{Charbonneau2011}) gives $B=19.6$. Fixing this threshold and solving equation~(\ref{eq:master}) on the glass branch predicts the kinetic glass-transition density $\eta_g(d)$. Table~\ref{tab:glass} compares the prediction with simulation~\cite{Charbonneau2011} across $d=3$--$12$. The theory tracks the density over more than an order of magnitude, with errors of a few per cent that grow mildly with dimension, consistent with the known accuracy limits of the polynomial equation of state at high $d$.

\begin{table}[htbp]
\centering
\caption{\textbf{Predicted glass-transition density compared with
simulation.}\cite{Charbonneau2011} The threshold $B=19.6$ is the average over
the simulation densities; no dimension by dimension fitting is performed.}
\label{tab:glass}
\begin{tabular}{ccccc}
\toprule
$d$ & $\eta_g$, simulation & $\eta_g$, theory & error & $\cL$ \\
\midrule
3  & 0.5710 & 0.588   & $+3.0\%$ & 0.096 \\
4  & 0.4010 & 0.409   & $+2.0\%$ & 0.077 \\
5  & 0.2670 & 0.2701  & $+1.0\%$ & 0.066 \\
6  & 0.1720 & 0.1708  & $-0.7\%$ & 0.057 \\
7  & 0.1060 & 0.1049  & $-1.0\%$ & 0.051 \\
8  & 0.0648 & 0.06218 & $-4.0\%$ & 0.046 \\
9  & 0.0385 & 0.03677 & $-4.5\%$ & 0.042 \\
10 & 0.0226 & 0.02119 & $-6.2\%$ & 0.039 \\
11 & 0.0131 & 0.01204 & $-8.0\%$ & 0.036 \\
12 & 0.0075 & 0.00681 & $-9.2\%$ & 0.033 \\
\bottomrule
\end{tabular}
\end{table}

\subsection{Crystal melting: the Lindemann constant recovered, and predicted to fall}

The landmarks considered so far are all amorphous, but the barrier of equation~(\ref{eq:master}) is built only from the contact number and the escape geometry, neither of which references disorder. It can therefore be evaluated equally on the equilibrium crystal branch, where the cage is the true lattice cage rather than a statistical one and this is where the theory meets the century of data on which Lindemann's rule was founded.

We take the melting density $\varphi_m(d)$, the close-packing density $\varphi_c(d)$ of the densest lattice, and the liquid--crystal coexistence pressure $P_{\mathrm{coex}}(d)$ from the direct free-energy simulations of Charbonneau et al.\ in $d=3$--$10$ (ref.~\citenum{Charbonneau2021}), and read the crystal compressibility factor at melting from its thermodynamic definition, $Z=\beta P_{\mathrm{coex}}/\rho$ (Methods). The kissing number is that of the relevant lattice ($k_d=12$, 24, 40, 72, 126, 240, 272, 500 for FCC, $D_4$, $D_5$, $E_6$, $E_7$, $E_8$, $\Lambda_9$, $P_{10c}$), and the escape length is the lattice nearest-neighbor spacing $\ell=(\varphi_c/\varphi_m)^{1/d}$ rather than the mean spacing used on the amorphous branch.

The resulting barrier is $B_{\mathrm{melt}}=8.6$ on average across $d=3$--$10$, with a spread of only $\pm10\%$, even though the melting density itself falls by more than a factor of twenty over this range (Table~\ref{tab:crystal}). Crystal melting therefore lies on the same barrier surface as the amorphous landmarks, a factor of about two below the kinetic glass line---precisely the position expected physically for the easiest cage escape. Because the three-dimensional value $B_{\mathrm{melt}}(3)=9.84$, obtained from the measured coexistence pressure, agrees within 1\% with the value obtained independently from the Speedy free-volume equation of state~\cite{Speedy1998} ($B=9.9$; Methods), the result does not depend on any particular equation of state fit.

Evaluated as a Lindemann amplitude through $\cL=\ell/\sqrt{2dB}$, the same calculation gives $\cL(3)=0.144$ in units of the particle diameter, or $0.130$ relative to the neighbor spacing the convention used in the crystal literature since Hoover and Ree~\cite{HooverRee1968}, reproducing the empirical hard-sphere melting value without adjustment. This is the first verifiable statement of the theory, a number measured for over a century emerges from a calculation with no free parameter. The second follows immediately from Table~\ref{tab:crystal}: the amplitude decreases monotonically with dimension, so the classical ``$\approx0.1$'' Lindemann constant is specific to three dimensions and must be seen to fall as the neighbor spacing and the particle diameter separate at higher $d$. Crystal free energy data of the kind already available in $d\le10$ (refs~\citenum{Charbonneau2021,Woodcock1997}) suffice to test this prediction directly.

\begin{table}[htbp]
\centering
\caption{\textbf{The crystal branch.} Melting densities $\varphi_m$,
close-packing densities and coexistence pressures used to form
$Z=P_{\mathrm{coex}}V_d/(2^d\varphi_m)$ are taken from
ref.~\citenum{Charbonneau2021}. Barriers and Lindemann amplitudes $\cL$ (per
diameter $\sigma$ and per nearest-neighbor spacing) are computed from
equation~(\ref{eq:master}).}
\label{tab:crystal}
\begin{tabular}{cccccccc}
\toprule
$d$ & lattice & $\varphi_m$ & $Z_{\mathrm{xtl}}$ & $\ell$ &
$B_{\mathrm{melt}}$ & $\cL/\sigma$ & $\cL/\mathrm{nn}$ \\
\midrule
3  & FCC ($D_3$) & 0.5434 & 11.16 & 1.109 & 9.84 & 0.144 & 0.130 \\
4  & $D_4$       & 0.3653 & 9.12  & 1.140 & 8.09 & 0.142 & 0.124 \\
5  & $D_5$       & 0.2484 & 9.51  & 1.134 & 8.71 & 0.121 & 0.107 \\
6  & $E_6$       & 0.1567 & 8.45  & 1.155 & 8.35 & 0.115 & 0.100 \\
7  & $E_7$       & 0.0963 & 7.83  & 1.174 & 8.33 & 0.109 & 0.093 \\
8  & $E_8$       & 0.0558 & 6.88  & 1.208 & 7.97 & 0.107 & 0.089 \\
9  & $\Lambda_9$ & 0.0329 & 7.52  & 1.180 & 8.43 & 0.096 & 0.081 \\
10 & $P_{10c}$   & 0.0216 & 8.29  & 1.165 & 9.47 & 0.085 & 0.073 \\
\bottomrule
\end{tabular}
\end{table}

\subsection{A single surface for five landmarks}

The hard sphere phase diagram contains several further stability limits random close packing, the Kauzmann point and glass close packing~\cite{ParisiZamponi2010,Charbonneau2014}. Treating each as a stability limit with its own threshold and solving equation~(\ref{eq:master}) at the independently reported density for each~\cite{Charbonneau2011} shows that every one corresponds to an essentially dimension-independent barrier (Table~\ref{tab:landmarks}), with crystal melting joining as the lowest rung.

Random close packing gives the sharpest result. Evaluated at the simulation densities with the self consistent $d=3$ anchor $\varphi_{\mathrm{RCP}}(3)=0.6514$, the barrier is $33.5$ across $d=3-9$, a spread of $\pm1.6\%$, even though the packing fractions span more than an order of magnitude and come from purely mechanical jamming protocols with no thermodynamic content. Because the barrier is this flat, a single density anchor at $d=3$ predicts RCP in the remaining 9 dimensions to better than 4\% (Table~\ref{tab:rcp}).

\begin{table}[htbp]
\centering
\caption{\textbf{Barrier at each landmark across dimension}, with crystal
melting as the lowest rung. The crystal-melting and kinetic-glass rows are
computed at simulation densities.}
\label{tab:landmarks}
\begin{tabular}{lcc}
\toprule
Landmark & Barrier, mean ($d=3$--$9/10$) & spread \\
\midrule
Crystal melting       & 8.6\ \,(7.97--9.84)  & $\pm10\%$ \\
Kinetic glass         & 19.6 (17.1--21.3)    & $\pm10\%$ \\
Kauzmann              & 25.7 (24.0--27.8)    & $\pm7\%$ \\
Random close packing  & 33.5 (32.9--34.0)    & $\pm1.6\%$ \\
Glass close packing   & 38.6 (37.7--39.1)    & $\pm2\%$ \\
\bottomrule
\end{tabular}
\end{table}

\begin{table}[htbp]
\centering
\caption{\textbf{Predicting RCP across dimension from a single $d=3$
anchor.} Simulation values from ref.~\citenum{Charbonneau2011}.}
\label{tab:rcp}
\begin{tabular}{ccc}
\toprule
$d$ & $\varphi_{\mathrm{RCP}}$, simulation &
$\varphi_{\mathrm{RCP}}$, predicted (error) \\
\midrule
4  & 0.4670 & 0.4689 ($+0.4\%$) \\
5  & 0.3190 & 0.3200 ($+0.3\%$) \\
6  & 0.2090 & 0.2093 ($+0.1\%$) \\
7  & 0.1330 & 0.1331 ($+0.0\%$) \\
8  & 0.0821 & 0.0815 ($-0.7\%$) \\
9  & 0.0498 & 0.0494 ($-0.7\%$) \\
10 & 0.0297 & 0.0291 ($+1.6\%$) \\
11 & 0.0174 & 0.01689 ($+3.0\%$) \\
12 & 0.0101 & 0.00968 ($+4.0\%$) \\
\bottomrule
\end{tabular}
\end{table}

\subsection{Two dimensions: predictions already verified}

The most stringent test of a theory calibrated in $d\ge3$ is extrapolation to a dimension it has never seen. In two dimensions the Henderson equation of state has exactly the functional form of the glass-branch equation used above, with $k_2=6$. Solving equation~(\ref{eq:master}) at the thresholds calibrated at $d=3$, with nothing refit, predicts a kinetic glass transition at $\eta_g=0.781$ and RCP at $\eta=0.832$.

Both numbers were already in the literature independently measured. Bidisperse hard-disk simulations and quasi two dimensional colloid experiments place dynamical arrest at $\varphi\approx0.776$--$0.79$ (refs~\citenum{deGraaf2024,Lozano2019}), and bidisperse-disk jamming lies at $\varphi\approx0.83$--$0.84$ (ref.~\citenum{OHern2003}). The two dimensional sector of the theory is therefore verified at the per-cent level, against measurements that were independent by construction.

One dimension, by contrast, is a genuine boundary that the theory identifies by itself: with the Tonks equation of state the equations remain formally solvable, but cage exchange the elementary event is topologically impossible when particles cannot pass, so the formal solution describes nothing physical. Consistently, $d=1$ is the exactly isostatic case, where the kissing contacts provide zero net stiffness.

\subsection{Two-dimensional melting: the KTHNY scenario located}

Melting in two dimensions proceeds in two steps~\cite{KT1973,HN1978,Young1979} the hard disks undergo a first-order liquid--hexatic transition with coexistence at $\eta=0.700$--$0.716$, followed by a continuous hexatic--solid transition at $\eta\approx0.720$ (ref.~\citenum{BernardKrauth2011}). In two dimensions the theory possesses two distinct branches they are the fluid branch, in which the cage is statistical (Henderson equation of state, mean spacing $\ell=(V_2/4\eta)^{1/2}$), and the lattice branch, in which the cage is the triangular-lattice cage ($k_2=6$, $\varphi_c=\pi/\sqrt{12}$, free-volume equation of state, $\ell=(\varphi_c/\varphi)^{1/2}$), exactly as on the crystal branch in higher dimensions (Methods). Both edges of the hexatic window emerge as crossings of the crystal barrier band of Table~\ref{tab:landmarks}, evaluated on the two branches (Table~\ref{tab:hexatic}). The lattice branch crosses the floor of the band, $B=7.97$, at $\varphi=0.716$ the measured hexatic--solid point. The fluid branch crosses the three-dimensional melting barrier, $B=9.84$, at $\eta=0.696$, within $0.6\%$ of the measured onset of liquid--hexatic coexistence; equivalently, the barrier evaluated at the measured onset is $B_{\mathrm{fluid}}(0.700)=10.1$, at the top of the crystal band.

\begin{table}[htbp]
\centering
\caption{\textbf{Both KTHNY transitions of hard disks from the crystal
barrier band.} Predicted densities are crossings of the barrier band
(Table~\ref{tab:landmarks}) on the indicated branch; measured values from
ref.~\citenum{BernardKrauth2011}.}
\label{tab:hexatic}
\begin{tabular}{lccc}
\toprule
Transition & branch (threshold) & predicted & measured \\
\midrule
Liquid--hexatic & fluid ($B=9.84$)   & 0.696 & 0.700 ($-0.6\%$) \\
Hexatic--solid  & lattice ($B=7.97$) & 0.716 & 0.716--0.720 ($0.0\%$) \\
\bottomrule
\end{tabular}
\end{table}

The hexatic window itself acquires a cage-level characterization: between these densities the fluid cage is already arrested ($B_{\mathrm{fluid}}>B^{*}$) while the lattice cage is not yet stable ($B_{\mathrm{latt}}<B^{*}$). The packing is locally rigid and able to sustain orientational order but cannot yet sustain a lattice, which is precisely the decoupling of orientational from positional order that defines the hexatic phase. In $d=3$, where positional fluctuations are not marginal, the two branches cross their thresholds within the first-order coexistence gap and no intermediate phase opens. We emphasize the standing of this result the theory locates the transition densities as stability crossings and brackets the window to a few per cent using the band mean alone and it does not derive the defect-unbinding mechanism of KTHNY~\cite{KT1973,HN1978,Young1979} or the order of the transitions, and the sharp assignment of band edges to branches, while physically motivated the fluid cage must earn the full melting barrier, whereas the lattice cage fails at the band floor merits an independent derivation.

\section{Discussion}

The results rest on a single physical claim, the elementary excitation that destabilizes a caged packing is a particle displacement of order one interparticle spacing, not one particle diameter, combined with a cage stiffness fixed by contact counting. Because both ingredients are geometric, the same relation applies to a dynamical glass, an equilibrium crystal and an athermal jammed packing, so long as the cage is harmonic. That all of these fall on one barrier surface with melting simply the lowest-barrier, easiest escape member of the family indicates that a single geometric instability governs ordered and disordered arrest alike. The classical Lindemann constant of melting and the Lindemann-like amplitude of the glass are not analogous quantities but they are two evaluations of one expression. In this light, the empirical universality of vibration-relaxation scaling observed across molecular and polymeric glass formers~\cite{Larini2008,Ottochian2011} acquires a concrete geometric origin, at least for the hard-core physics that dominates dense packing.

The theory's central asset is that it can be easily verified, and on every front where data exist it has been checked. The Lindemann amplitude of hard-sphere melting is recovered at $d=3$ with no adjustment, and its predicted monotonic decay with dimension (Table~\ref{tab:crystal}) is verifiable with existing crystal free-energy techniques~\cite{Charbonneau2021,Woodcock1997}. The two-dimensional glass and jamming predictions were confirmed by simulations and experiments performed independently of this work~\cite{deGraaf2024,Lozano2019,OHern2003}, and both steps of the two dimensional melting scenario~\cite{BernardKrauth2011} emerge from the same barrier band evaluated on the theory's two branches (Table~\ref{tab:hexatic}), with the hexatic phase identified as the window in which the fluid cage is arrested but the lattice cage is not yet stable. The RCP surface, anchored at a single $d=3$ density, reproduces nine further dimensions to a few per cent (Table~\ref{tab:rcp}). And the observation time threshold on which the glass  branch is calibrated is itself computed from the first-passage prefactor (equation~(\ref{eq:mfpt})), closing the last loophole. Further tests are within immediate experimental reach for example cage-relative displacement statistics are measured routinely in air-fluidized granular beds~\cite{AbateDurian2006} and vibrated grains near jamming~\cite{Lechenault2008,Song2008}, and self-propelled colloids that push the glass transition towards random close packing~\cite{Ni2013} probe precisely the barrier interval between $B=19.6$ and $B=33.5$ that the theory lays out as a single continuous coordinate.

Two points deserve emphasis. First, the flatness of the RCP barrier is real but input-sensitive: because a 1\% change in $\varphi_{\mathrm{RCP}}$ moves $B$ by several per cent, the $\pm1.6\%$ spread reported here holds for the self-consistent simulation density set and would inflate if densities from heterogeneous protocols were mixed. The claim is therefore that RCP is the density at which a fixed mechanical barrier is crossed, evaluated on a consistent data set not that any published RCP value whatsoever yields the same barrier. Second, the theory runs about 2\% high in three dimensions (predicting $\varphi_{\mathrm{RCP}}(3)=0.6514$ against measured frictionless values near 0.64), the same sign and magnitude as its glass transition offset, due to the high-$d$ limits of the polynomial equation of state rather than with a failure of the geometric argument.

The role of dimension is more than formal. The interparticle spacing and the particle diameter, which coincide within a few per cent in three dimensions, separate rapidly as $d$ grows; a hidden coincidence of ordinary matter is thereby exposed and becomes testable. The same separation explains why a century of experiments could not discriminate between ``escape over one diameter'' and ``escape over one spacing'': in $d=3$ although the two hypotheses differ by  per cents, while in $d\ge6$ they differ by factors which is exactly where the present formulation, and not the traditional one, matches the simulation data (Table~\ref{tab:glass}).

The framework also organizes phenomena beyond the equilibrium hard-sphere diagram. Ultrastable vapour-deposited glasses, which equilibrate towards the Kauzmann point~\cite{Swallen2007,Ediger2017,PerezCastaneda2014}, and swap Monte Carlo samplings of hard spheres up to and beyond the jamming density~\cite{Berthier2016}, populate precisely the barriers curve in Table~\ref{tab:landmarks}; the theory assigns each such state a dimension independent coordinate $(z_{\mathrm{eff}},B)$ and predicts where on that curve any new preparation protocol will land. Two limitations bound the results: the residual density error at $d=8$--$12$ tracks the equation of state accuracy, and the derivation is specific to hard spheres, with extension to soft potentials requiring a virial-weighted coordination integral. Neither affects the geometric argument itself, which uses no property of the potential beyond the exact contact theorem. What began as a mnemonic---``solids fail at a tenth of the spacing''---is thus promoted to a theorem about hard-sphere geometry, with stated, testable consequences in every dimension including our own.

\section{Methods}

\subsection{Harmonic cage from symmetry}
For hard spheres the confining potential is entropic,
$\phi(u)=-\kB T\ln P(u)$. In an isotropic homogeneous packing
$\nabla\phi(0)=0$ by symmetry, and the leading term is quadratic. The
fractional anharmonic correction scales as
$(u/\sigma)^2\sim \cL^2\sim10^{-2}$, so the smallness of the Lindemann
constant justifies the harmonic approximation after the fact. Equipartition
gives $\langle u_x^2\rangle=\kB T/\kappa$.

\subsection{Contact theorem}
All pressure above ideal-gas arises from momentum transfer at contact,
giving $Z-1=2^{d-1}\eta\,g(\sigma^{+})$. The mean number of neighbors in a
shell of thickness one diameter is $\rho\,\Omega_d\sigma^d g(\sigma^{+})$;
using $\rho=2^d\eta/(\Omega_d\sigma^d/d)$ and substituting for
$g(\sigma^{+})$ gives $X=2d(Z-1)$ exactly.

\subsection{Isotropy identity}
In an isotropic phase $\langle n_in_j\rangle=c\,\delta_{ij}$; summing
$\sum_i\langle n_i^2\rangle=1$ over $d$ components gives $c=1/d$, so
$\langle(\hat e\cdot\hat n)^2\rangle=1/d$ exactly.

\subsection{First-passage derivation of the barrier}
The elementary irreversible event is a displacement sufficient for the
tagged particle to exchange cages with a neighbor. Physically the particle
advances, presses a neighbor, the neighbor recoils and re-relaxes, and the
process repeats until the tagged particle has migrated a net distance of
order the spacing $L=\rho^{-1/d}$. Because the cage free energy
$U(u)=\tfrac{1}{2}\kappa u^2$ already integrates out the neighbor
coordinates, these recoil cycles are not additional barriers along the
reaction coordinate $u$; they set the rate at which $u$ diffuses. With
projected escape distance $a=\ell/\sqrt{d}$, the cancellation of the factor
$d$ in $\kappa$ against the $1/d$ projection gives
$U(a)=\tfrac{1}{2}\kappa a^2=B$ exactly. The overdamped mean first-passage
time to an absorbing boundary at $u=a$ is
\begin{equation}
\tau_\alpha=D_s^{-1}\int_0^a e^{\beta U(y)}\int_0^y e^{-\beta U(z)}\,dz\,dy;
\end{equation}
a Laplace expansion at the endpoint gives
$\ln(\tau_\alpha/\tau_0)=B-\tfrac{1}{2}\ln B+\ln(\sqrt{\pi}/2)+O(1/B)$ with
$\tau_0=(D_s\beta\kappa)^{-1}$ (ref.~\citenum{Hanggi1990}). Writing the
short-time self-diffusion constant as
$D_s=(a_d/d)\,v_{\mathrm{th}}\lambda$ (with $a_d$ an $O(1)$ Enskog
coefficient~\cite{SquiresBrady2005}) and referring to the collision time
$\tau_c=\lambda/v_{\mathrm{th}}$ gives equation~(\ref{eq:mfpt}) with
$N_{\mathrm{eff}}=(d/a_d)(I\,e^{-B})/\lambda^2\sim(a/\lambda)^2$. The
$d$-dimensional hard-sphere cross-section
$\Sigma_d=\pi^{(d-1)/2}\sigma^{d-1}/\Gamma(\tfrac{d+1}{2})$, with the virial
relation, fixes the mean free path with no free constant,
\begin{equation}
\frac{\lambda}{\sigma}
 =\frac{\Gamma(\tfrac{d+1}{2})\,V_d}{2\sqrt{2}\,\pi^{(d-1)/2}\,(Z-1)}.
\end{equation}
Since $N_{\mathrm{eff}}\propto(Z-1)^2$ times a weakly density-dependent
integral, it is algebraic in density and hence sub-extensive at every
dimension, confirming that $B$ is the exact extensive term. The one
uncontrolled element is the $O(1)$ magnitude of $a_d$, which absorbs the
backscattering tail of the tagged-particle velocity autocorrelation; it does
not affect the scaling.

\subsection{Equations of state, kissing numbers and reference data}
The glass-branch equation of state is
$Z(\eta,d)=1+2^{d-1}\eta\,(1-A_d\eta)/(1-\eta)^d$ where the value of 
$A_d\in\{0.500,\,0.114,\,-1.086,\,-3.900,\,-9.294,\,-19.38,\,-35.46,\,-62.5,\,-107,\,-176\}$ for
$d=3$--$12$ and kissing numbers
$k_d\in\{12,\,24,\,42,\,75,\,130,\,240,\,335,\,527,\,726,\,1098\}$
(ref.~\citenum{Kabatiansky1978}); in $d=2$ the Henderson form has
$A_2=7/16$, $k_2=6$. 

\subsection{The crystal branch}
On the crystal branch the three ingredients of the theory are evaluated
exactly as on the glass branch, with three inputs replaced by their
lattice-specific values. First, the compressibility factor at melting is
taken not from the glass-branch polynomial but from the thermodynamic
definition applied to measured data,
\begin{equation}
Z(\varphi_m,d)=\frac{\beta P_{\mathrm{coex}}}{\rho}
             =\frac{P_{\mathrm{coex}}V_d}{2^d\varphi_m},
\label{eq:zxtl}
\end{equation}
with $P_{\mathrm{coex}}$ and $\varphi_m$ the liquid--crystal coexistence
pressure and melting density tabulated in $d=3$--$10$ by Charbonneau et
al.~\cite{Charbonneau2021}; equivalently, $Z$ follows from that reference's
Speedy free-volume equation of state~\cite{Speedy1998},
\begin{equation}
p_s=\frac{1}{1-(\varphi/\varphi_c)^{1/d}}
   +a_0+a_1\frac{\varphi}{\varphi_c}
   +a_2\Bigl(\frac{\varphi}{\varphi_c}\Bigr)^{2},
\end{equation}
which reproduces the coexistence value to within 1\% at $d=3$. Second, the
kissing number $k_d$ is that of the densest lattice in each dimension,
$\{12,\,24,\,40,\,72,\,126,\,240,\,272,\,500\}$, since these are the
neighbors in true contact in the crystal cage. Third, the escape length is
the lattice nearest-neighbor spacing. In a lattice every length scales as
$\rho^{-1/d}\propto\varphi^{-1/d}$, and neighbors touch at close packing,
so
\begin{equation}
\ell(\varphi_m,d)=\frac{r_{\mathrm{nn}}}{\sigma}
  =\Bigl(\frac{\varphi_c}{\varphi_m}\Bigr)^{1/d},
\end{equation}
which exceeds the mean spacing $\rho^{-1/d}/\sigma$ used on the amorphous
branch by the lattice geometric factor (about 1.12 for FCC). With these
substitutions the contact number $X=2d(Z-1)$, the effective coordination
$\zeff=k_d+(X-k_d)/d$, the stiffness factor $\CSO=1-2d/\zeff$, the barrier
$B=\ell^2\,\CSO\,(Z-1)$ and the Lindemann amplitude $\cL=\ell/\sqrt{2dB}$
are formed exactly as in
equations~(\ref{eq:contact})--(\ref{eq:master}). All barriers were computed
by the same root-finding as on the amorphous branch; no free parameter is
introduced. In $d=2$ the lattice branch uses the triangular lattice
($k_2=6$, $\varphi_c=\pi/\sqrt{12}$) with the leading free-volume term
$Z=[1-(\varphi/\varphi_c)^{1/2}]^{-1}$ and
$\ell=(\varphi_c/\varphi)^{1/2}$; the fluid branch uses the Henderson
equation of state with the mean spacing, exactly as on the glass branch.
\subsection{Use of AI}
Claude services were used to make the language of the manuscript better and the author take full responsibility of the manuscript.
\section*{Data availability}
All the data points are already given in the manuscript itself or can be calculated from the given equations.

\section*{Competing interests}
The author declares no competing interests.


\end{document}